\documentclass{article}
\usepackage{dcase2022,amsmath,graphicx,url,times,booktabs, tabularx}

\usepackage{multirow}
\usepackage{hyperref}

\title{Is my automatic audio captioning system so bad?\\SPIDE\MakeLowercase{r-max}: a metric to consider several caption candidates}

\name{Etienne Labbé, Thomas Pellegrini, Julien Pinquier}
\address{IRIT, Université Paul Sabatier, CNRS, Toulouse, France \\
    \{etienne.labbe, thomas.pellegrini, julien.pinquier\}@irit.fr \\
}

\begin{document}

\ninept
\maketitle

\begin{sloppy}

\begin{abstract}
Automatic Audio Captioning (AAC) is the task that aims to describe an audio signal using natural language. AAC systems take as input an audio signal and output a free-form text sentence, called a caption. Evaluating such systems is not trivial, since there are many ways to express the same idea. For this reason, several complementary metrics, such as BLEU, CIDEr, SPICE and SPIDEr, are used to compare a single automatic caption to one or several captions of reference, produced by a human annotator. Nevertheless, an automatic system can produce several caption candidates, either using some randomness in the sentence generation process, or by considering the various competing hypothesized captions during decoding with beam-search, for instance. If we consider an end-user of an AAC system, presenting several captions instead of a single one seems relevant to provide some diversity, similarly to information retrieval systems. In this work, we explore the possibility to consider several predicted captions in the evaluation process instead of one. For this purpose, we propose SPIDEr-max, a metric that takes the maximum SPIDEr value among the scores of several caption candidates. To advocate for our metric, we report experiments on Clotho v2.1 and AudioCaps, with a transformed-based system. On AudioCaps for example, this system reached a SPIDEr-max value (with 5 candidates) close to the SPIDEr human score of reference.
\end{abstract}


\section{Introduction}
\label{sec:intro}
Automated Audio Captioning (AAC) is the task, in which a system takes an audio signal as input and provides a short description of its content using natural language. 
AAC could be useful for hearing-impaired people, in machine-to-machine interaction, surveillance and information retrieval in general. In the last few years, the research community has developed a keen interest in AAC, in particular thanks to the Detection and Classification of Acoustic Scenes and Events (DCASE) Challenges and Workshops~\footnote{\url{http://dcase.community/}}, which have provided datasets and benchmarks for this task.


Most AAC systems use deep neural networks with a sequence-to-sequence encoder-decoder architecture, to build a semantic audio representation and generate a valid sentence as output~\cite{xu2022comprehensive}. They rely on models pretrained on large-scale datasets, to solve the data scarcity issue in AAC~\cite{kong_panns_2020,xinhao2021_t6,gontier:hal-03522488}.


In this work, we are interested in the evaluation of AAC systems. AAC evaluation borrows metrics from machine translation and image captioning, and consists of comparing a candidate caption to one or several manually produced captions of reference. Since evaluating text generated automatically is a difficult problem, several metrics are used in combination.
We investigate in particular the SPIDEr metric~\cite{liu_improved_2017}, a short name used to designate the average of two metrics called Consensus-based Image Description Evaluation (CIDEr)~\cite{vedantam_cider_2015} and Semantic Propositional Image Caption Evaluation (SPICE)~\cite{anderson_spice_2016}. SPIDEr is used, for instance, in the DCASE yearly challenges to rank the participant AAC systems~\footnote{\url{http://dcase.community/challenge2022/}}. 

In this paper, we report experiments using the AAC system we developed to participate in the DCASE 2022 AAC task. Like most AAC systems, we use a beam search decoder that allows to generate several candidate captions. The most likely one is used to compute the SPIDEr score of our system. 
A strong limitation of SPIDEr, in our opinion, is that only one caption candidate is considered for evaluation. As we shall illustrate in this paper, two correct captions that differ by a single word may have very different SPIDEr scores, if one of the words happens to be in the caption(s) of reference. To overcome this issue, we propose a metric that we call SPIDEr-max, which takes into account multiple candidates for a single audio recording. 

\section{Metrics}
\label{sec:metrics}

In the literature, most AAC systems are evaluated using the CIDEr, SPICE or SPIDEr metrics. These metrics come from the field of image captioning and evaluate a single candidate caption against a reference set. 

\subsection{CIDEr}
CIDEr~\cite{vedantam_cider_2015} is a metric based on the TF-IDF (term frequency-inverse document frequency) scores of each n-gram of the candidate and reference sentences. TF-IDF is used to give a higher weight to infrequent n-grams and lower weight to frequent n-grams.

The CIDEr metric calculation starts by stemming all the words and compute all the n-grams of size 1 to $N$ across all candidates and references. The frequency of each n-gram in references are used to compute TF-IDF of all captions. This means that the score of each candidate does not only depend on its corresponding references, but also on all the other references of the corpus being evaluated. Then, the TF-IDF scores are vectorized and used to compute cosine similarity between the candidate and each reference. The similarities are rescaled by a factor of 10 and averaged across the references to get the final score of the candidate. All the scores are averaged again to get the global score on a dataset.

The CIDEr-D metric is a more robust version of CIDEr supposed to be closer to human judgement. It removes the stemming operation to take into account the tense and plural of words, adds a penalty factor, and limits the maximum occurrence of candidate n-grams to penalize longer repetitive sentences. The penalty is multiplied by a similarity measure based on the length of the candidate $c$ and the reference $r$:
\begin{equation}
    \label{eq:cider_penalty}
    \text{Penalty}(c, r) = \exp \Big(-\frac{(|c| - |r|)^2}{2 \sigma^2} \Big)
\end{equation}

Some AAC papers do not specify which version of CIDEr they use, but in this paper we report CIDEr-D scores as used in the DCASE challenge. We use the default settings of CIDEr-D with the maximum n-gram size $N$ set to 4, and the hyperparameter $\sigma$ used for the penalty set to 6.


\subsection{SPICE}
SPICE~\cite{anderson_spice_2016} attempts to extract the semantic content of a sentence. Sentences are used as input to a Probabilistic Context-Free Grammar dependency parser\cite{klein_accurate_2003}, with several additional rules to build a dependency tree where each node is a word and each edge is a syntactic dependency. Custom rules are used to compute another graph, a ``semantic scene graph'', comprised of three types of nodes: objects, attributes and relations. Attributes are linked to a single object, and relations connect objects between them. The reference graphs are merged into one to be compared with a candidate graph. Then, the scene graphs are converted into lists of word tuples. An object is a tuple with the object name, an attribute is a tuple of two words with the object and attribute names, and a relation is a tuple of three words containing the two objects connected and the relation names. Finally, the list is binarized for the candidate and the references, and used to compute an F-Score.

The M-SPICE metric~\cite{hsu_text-free_2020} is a variant of SPICE, which takes multiple candidates for a single audio. This metric was introduced to evaluate the diversity of the words used in multiple candidates generated by stochastic decoding methods. The only difference is that the semantic graph of each candidate is merged into one, exactly as for the reference list. The other steps remain the same.


        

\subsection{SPIDEr}
SPIDEr~\cite{liu_improved_2017} is a metric originally used as a cost function to optimize a model on SPICE and CIDEr-D at the same time. SPIDEr is the average of CIDEr-D and SPICE, and is supposed to have the benefits of both previous metrics. Since CIDEr-D gives a score between 0 and 10 and SPICE between 0 and 1, the SPIDEr score is between 0 and 5.5, which is quite uncommon for a metric. SPIDEr is usually the metric used in AAC papers to compare models, even if other machine-translation metrics like BLEU~\cite{papineni_bleu_2001}, ROUGE-L~\cite{lin-2004-rouge}, and METEOR~\cite{denkowski_meteor_2014} scores are also reported.

\section{System description}
\label{sec:system_description}

\subsection{Datasets}
\label{ssec:datasets}

The Clotho v2.1 dataset~\cite{drossos_clotho_2019} is an audio captioning dataset containing 6974 audio files of approximately 43.6 hours from Freesound between 15 and 30 seconds. Each audio is described by 5 captions annotated by humans. The dataset is divided in 3 different splits: development, validation and evaluation, which corresponds to development-training, development-validation and development-testing, the conventional names used in the DCASE Challenge. In this paper, we use these names. The training subset contains 217362 words with a caption length between 8 and 20 words.

AudioCaps~\cite{kim-etal-2019-audiocaps} is another audio captioning dataset containing 49838 training files of approximately 136.6 hours from AudioSet~\cite{gemmeke_audio_2017}, a large audio tagging dataset with audio extracted from YouTube. AudioCaps contains only 1 caption per audio in the training subset and 5 captions for the validation and testing subsets. Since YouTube removes videos uploaded by users for various reasons, our version of AudioCaps contains only 46230 over 49838 files in training subset, 464 over 495 in validation subset and 912 over 975 files in testing subset. Our training subset contains 402482 words with a caption length between 1 and 52 words.

To extract audio features, we resample audio signals to 32 kHz and compute log-Mel spectrograms with a window size of 32 ms, a hop size of 10 ms and 64 Mel bands. All captions are put in lowercase and punctuation characters are removed. We used the spaCy tokenizer~\cite{ines_montani_2021_5764736} to split sentences into words, resulting in a vocabulary of 4370 tokens for Clotho and 4724 words for AudioCaps.

\subsection{Model architecture}
\label{ssec:model_architecture}
We adopt a standard encoder-decoder structure used in most AAC systems, with a pre-trained encoder to extract audio features and a transformer decoder to generate our captions. The encoder is the CNN10 model, a convolutional network from the Pretrained Audio Neural Networks study (PANN)~\cite{kong_panns_2020}. We used the weights available on Zenodo\footnote{\url{https://zenodo.org/record/3987831}} to initialize the model at the beginning of the training. An affine layer was added to project 512-dimensional to 256-dimensional embeddings. We kept the time axis of the audio embedding used as input for the decoder.

The decoder is a standard transformer decoder~\cite{DBLP:journals/corr/VaswaniSPUJGKP17}. It takes the audio embeddings as inputs and all the previous words predicted. The word embeddings are randomly initialized and learned during training. We use teacher forcing with cross-entropy to train the model. During the testing phase, captions are generated using beam search, and we select the best candidate using the probability of the sentence $P$ given by the model. The combination of our encoder and decoder is simply named ``CNN10-Transformer''.

\subsection{Experimental setup}
\label{ssec:experimental_setup}
We trained models for 50 epochs, on both datasets separately. To optimize our networks, we used Adam~\cite{kingma_adam_2017}, with a learning rate set to $5.10^{-4}$ at the first epoch, a $10^{-6}$ weight decay, a 0.9 $\beta_1$ and 0.999 $\beta_2$, and $\epsilon$ set to $10^{-8}$. We used a cosine learning rate scheduler with the following rule:
\begin{equation}
    \text{lr}_k = \frac{1}{2} \bigg(1 + \cos \Big( \frac{k \pi}{K} \Big) \bigg) \text{lr}_0
\end{equation}
with $k$ being the current epoch index, and $K$ the total number of epochs.

The transformer decoder uses an embedding dimension $d_{model}$ of 256, four attention heads $h$, six stacked standard decoder layers, and a global dropout $P_{drop}$ set to 0.2. The last affine layer projects the 256-dimensional embeddings to an output of the vocabulary size of the dataset. We used label smoothing to reduce overfitting, set to 0.1 for AudioCaps and 0.2 for Clotho. In order to avoid gradient explosion, we clip gradients by a maximal L2-norm value set to 10 and 1 for AudioCaps and Clotho, respectively. During testing,  beam size is set to 8 for Clotho and 2 for AudioCaps.
The final encoder-decoder model results in 16M trainable parameters. We also used SpecAugment~\cite{park19e_interspeech} as audio data augmentation with two bands dropped on the time axis with a maximal size of 64 bins and one band dropped on the frequency axis with a maximum size of two bins. Our implementation uses PyTorch~\cite{NIPS2019_9015}, PyTorch-Lightning~\cite{falcon2019pytorch} and our aac-datasets~\footnote{\url{https://pypi.org/project/aac-datasets}} package to download and manage audio captioning datasets.

\section{\texorpdfstring{SPIDE\MakeLowercase{r}}{SPIDEr} results}
\label{sec:spider_results}
Results on Clotho and AudioCaps of our model CNN10-Transformer are shown in Table~\ref{table:spider_results_baseline}. Standard deviations of our model are very small (0.001 and 0.004 for Clotho and AudioCaps, respectively). Cross-reference scores are computed by using one of the reference as a candidate and the four others as references five times.

\begin{table}[!htb]
    \centering
    \caption{SPIDEr scores on Clotho v2.1 and AudioCaps with state-of-the-art results and cross-reference scores.}
    \label{table:spider_results_baseline}
    \vspace{0.2cm}
    \begin{tabular}{@{}lcc@{}}
        \toprule
        \textbf{System}   & \textbf{Clotho} & \textbf{AudioCaps} \\ \midrule
        Best (SOTA) & 0.320~\cite{xu2022_t6a} & 0.465~\hphantom{a}\cite{gontier:hal-03522488} \\
        Human & N/A\hphantom{abcd} & 0.565\hphantom{}~\cite{kim-etal-2019-audiocaps} \\
        Cross-Referencing & 0.573\hphantom{abcd} & 0.564\hphantom{abcd} \\ \midrule
        CNN10-Transformer (ours) & 0.247\hphantom{abcd} & 0.401\hphantom{abcd} \\
        \bottomrule
    \end{tabular}
\end{table}


Our model performs much better on AudioCaps than Clotho, with a SPIDEr score of 0.401 and 0.247, respectively. It is also closer to the cross-reference and human scores in AudioCaps. This is probably due to the fact that the CNN10 encoder has been pre-trained on AudioSet, which is a superset of AudioCaps. In addition, the captions in AudioCaps are simpler than those in Clotho, with shorter sentences and a relatively smaller vocabulary. The current best score on AudioCaps is also much closer to the cross-reference top score (0.100 difference) than the one on Clotho (0.253 difference).

\section{\texorpdfstring{SPIDE\MakeLowercase{r}}{SPIDEr} limitations}
\label{sec:spider_limitations}

\subsection{The SPIDEr score varies greatly between beam search candidates}
Tables~\ref{table:exemple_1} and \ref{table:exemple_2} show examples of candidates and captions of reference, one from Clotho, one from AudioCaps. The probability $P$ given by the model is also indicated. It is used to select the best candidate among the beam search hypotheses. We also reported the SPIDEr score associated to each candidate.

In the Clotho example, the most likely caption candidate is also the one with the highest SPIDEr score, based on the fact that the rather rare word ``tin'' was found by the automatic system. Thus, in this example, the differences observed between the various hypotheses seem justified. On the contrary, in the second example, from AudioCaps, the most likely automatic caption is different from the one with the highest SPIDEr score.    

\begin{table}[!htb]
    \footnotesize
    \centering
    \caption{Captions for the Clotho development-testing file named ``rain.wav''.}
    \label{table:exemple_1}
    \vspace{0.2cm}
    \begin{tabular}{@{}p{5.5cm}cc@{}}
        \toprule
        \textbf{Candidates} & \textbf{P} & \textbf{SPIDEr} \\ \midrule
        heavy rain is falling on a roof & 0.361 & 0.562 \\
        heavy rain is falling on \textbf{a tin roof} & \textbf{0.408} & \textbf{0.930} \\
        a heavy rain is falling on a roof & 0.369 & 0.594 \\
        a heavy rain is falling on the ground & 0.351 & 0.335 \\
        a heavy rain is falling on the roof & 0.340 & 0.594 \\ \midrule
        \multicolumn{3}{@{}l@{}}{\textbf{References}} \\ \midrule
        \multicolumn{3}{@{}l@{}}{heavy rain falls loudly onto a structure with a thin roof} \\
        \multicolumn{3}{@{}l@{}}{heavy rainfall falling onto a thin structure with a thin roof} \\
        \multicolumn{3}{@{}l@{}}{it is raining hard and the rain hits \textbf{a tin roof}} \\
        \multicolumn{3}{@{}l@{}}{rain that is pouring down very hard outside} \\
        \multicolumn{3}{@{}l@{}}{the hard rain is noisy as it hits \textbf{a tin roof}} \\ \bottomrule
    \end{tabular}
\end{table}


\begin{table}[!htb]
    \footnotesize
    \centering
    \caption{Captions for an AudioCaps testing file (id: \mbox{`jid4t-FzUn0'}).}
    \label{table:exemple_2}
    \vspace{0.2cm}
    \begin{tabular}{@{}p{5.5cm}cc@{}}
        \toprule
        \textbf{Candidates} & \textbf{P} & \textbf{SPIDEr} \\ \midrule
        a woman speaks and a sheep bleats & 0.475 & 0.190 \\
        a woman \textbf{speaks and a goat bleats} & 0.464 & \textbf{1.259} \\
        a man speaks and a sheep bleats & 0.464 & 0.344 \\
        an adult male speaks and a sheep bleats & 0.450 & 0.231 \\
        an adult male is speaking and a sheep bleats & \textbf{0.491} & 0.189 \\ \midrule
        \multicolumn{3}{@{}l@{}}{\textbf{References}} \\ \midrule
        \multicolumn{3}{@{}l@{}}{a man speaking and laughing followed by a goat bleat} \\
        \multicolumn{3}{@{}l@{}}{a man is speaking in high tone while a goat is bleating one time} \\
        \multicolumn{3}{@{}l@{}}{a man speaks followed by a goat bleat} \\
        \multicolumn{3}{@{}l@{}}{a person \textbf{speaks and a goat bleats}} \\
        \multicolumn{3}{@{}l@{}}{a man is talking and snickering followed by a goat bleating} \\ \bottomrule
    \end{tabular}
\end{table}

The agreement accuracy between the best candidate according either to the likelihood and to the SPIDEr score is only of 26.5\% on Clotho, and 22.6\% on AudioCaps. The correlation coefficient on all the likelihoods and the SPIDEr scores is 0.224 on Clotho and 0.259 on AudioCaps. This shows that the maximum candidate likelihood $P$ does not select the best caption according to the SPIDEr score.

\subsection{Can we choose a better candidate automatically?}
Selecting automatically the best candidate among the beam search hypotheses is a difficult problem: most candidates are very similar and usually describe the same events with different words. Figure~\ref{fig:audiocaps_beam_indexes} shows the histogram of the beam hypothesis indices that give the maximum SPIDEr score possible, for each candidate list when using a beam size of five. It reveals that no beam index seems better than another \textit{a priori}. The same conclusion can be drawn with Clotho. We tried to automatically select the best candidate using several features: vocabulary size, sentence length, and even with a shallow neural network trained to rank the sentences, but all these approaches failed to significantly improve the global SPIDEr score. 

\begin{figure}[!htb]
    \centering
    \includegraphics[width = 0.8\linewidth]{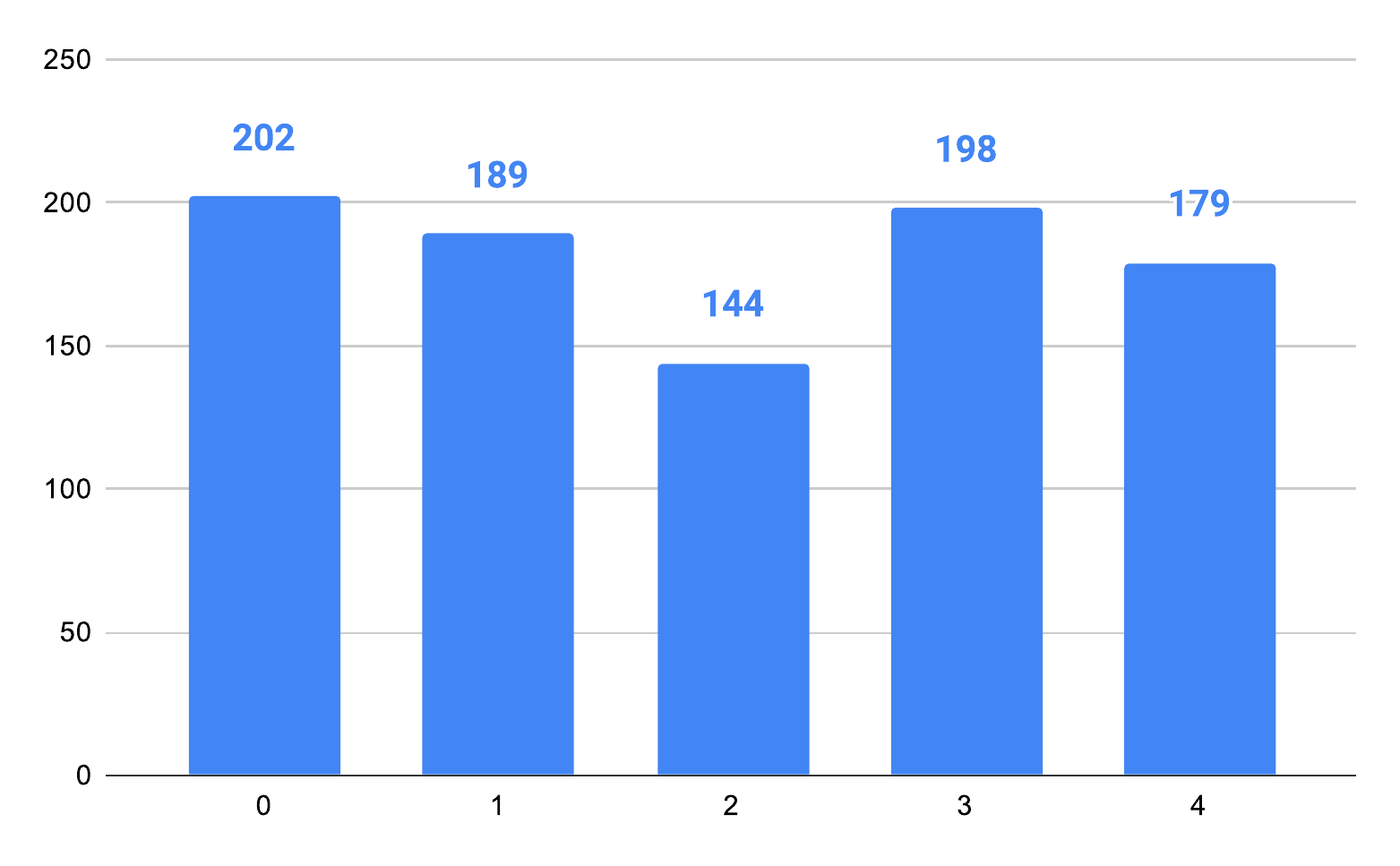}
    \caption{SPIDEr best beam indexes on AudioCaps testing subset}
    \label{fig:audiocaps_beam_indexes}
\end{figure}

To overcome these limitations, we propose to consider all the candidates produced by the model and select the best SPIDEr score between them with a new metric.



\section{\texorpdfstring{SPIDE\MakeLowercase{r-max}}{SPIDEr-max}}
\label{sec:spider-max}

\subsection{Definition}
We propose SPIDEr-max, defined by the following equation:
\begin{equation}
    \text{SPIDEr-max}(C, R) = \max_i \text{ SPIDEr}(C_i, R)
\end{equation}
where $C$ is a list of $N$ caption candidates and $R$ a list of references.

It consists of retaining the largest SPIDEr score among the scores calculated for a set of caption candidates, to avoid having to choose a single hypothesis. The SPIDEr-max values are between 0 and 5.5, like the SPIDEr score. The source code in PyTorch will be made available on GitHub\footnote{\url{https://github.com/Labbeti/spider-max}} upon paper acceptance.

\subsection{Results}
The score of SPIDEr-max highly depends on how many candidates we use, so we report results with various beam sizes in figures~\ref{fig:audiocaps_spider_max} and~\ref{fig:clotho_spider_max} for AudioCaps and Clotho, respectively. We varied the beam size from 1 to 10. If we imagine a human end-user, proposing at most 5 candidates captions would be reasonable, in our opinion.

\begin{figure}[!htb]
    \centering
    \includegraphics[width=1.0\linewidth]{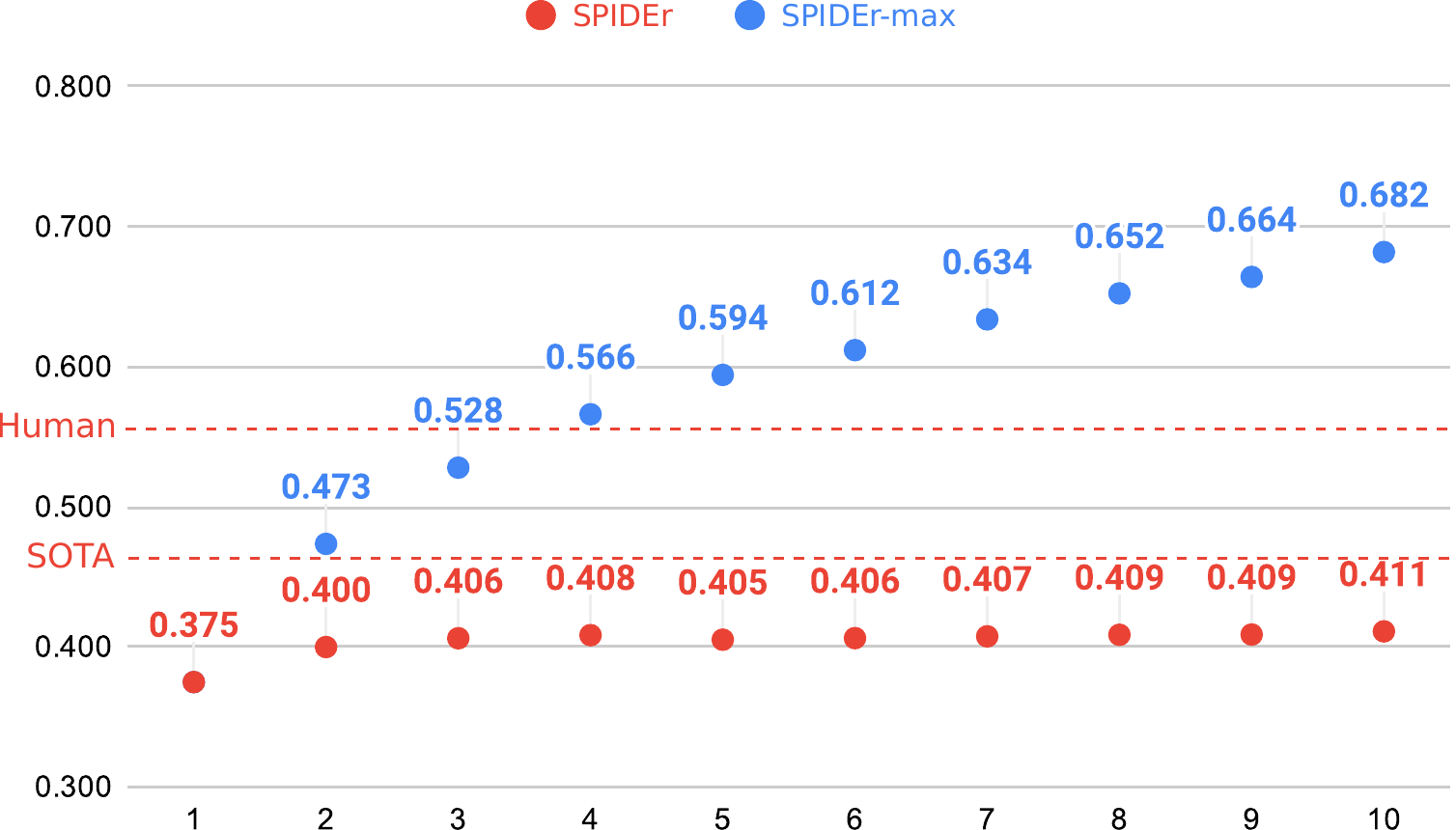}
    \caption{SPIDEr and SPIDEr-max scores with different beam sizes, calculated on the AudioCaps testing subset with CNN10-Transformer.}
    \label{fig:audiocaps_spider_max}
\end{figure}

\begin{figure}[!htb]
    \centering
    \includegraphics[width=1.0\linewidth]{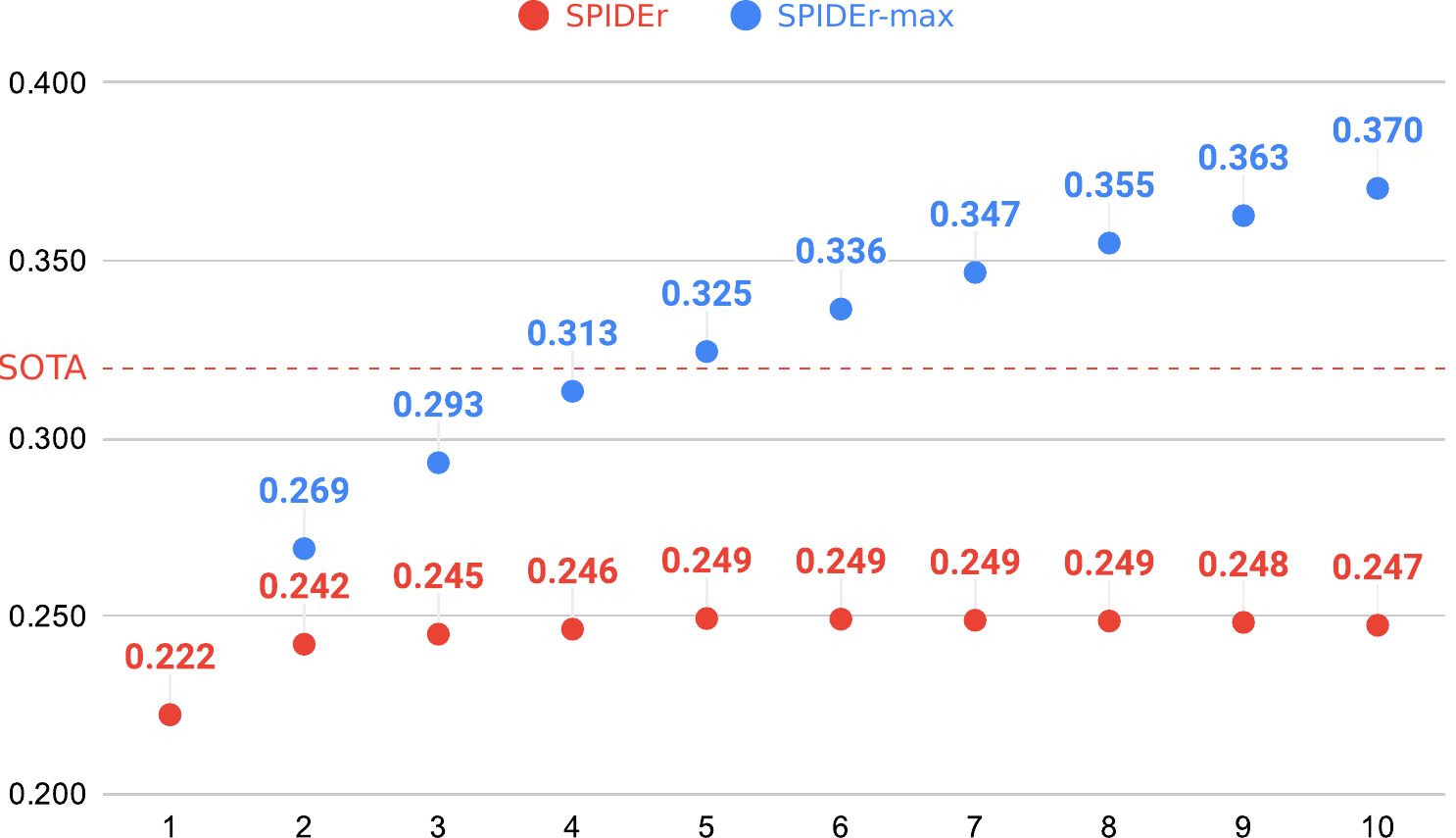}
    \caption{SPIDEr and SPIDEr-max scores with different beam sizes, calculated on the Clotho development-testing subset with CNN10-Transformer.}
    \label{fig:clotho_spider_max}
\end{figure}

On AudioCaps, the SPIDEr-max score increases rapidly above the score of our model from 0.401 to 0.473 with only a beam size of two. The scores continue to rise above the human SPIDEr score (0.565), meaning that our model is already producing human-like captions, but fail to select them, if we take the maximum likelihood criterion.


On Clotho, the scores also increase with a higher beam size, but they do not reach the cross-reference score on the first beam sizes. This is probably due to the references of Clotho, which show more diversity in terms of vocabulary and n-grams than AudioCaps.

We also tried to compute the SPIDEr-max score for a beam size equal to 100, which gave 0.953 on AudioCaps and 0.535 on Clotho, but we decided to focus on a few candidates, as it would be more realistic in a real scenario, where, for instance, automatic captions are proposed to an end-user.

\subsection{Why such a boost in SPIDEr-max?}
As we saw in the previous section, SPIDEr-max increases rapidly and even outreaches the human SPIDEr score on AudioCaps. We also noticed that predicting a correct infrequent n-gram seems to drastically improves the score of a candidate, probably due to the CIDEr-D metric based on the TF-IDF of the n-grams. To see if there is a relation between TF-IDF and the SPIDEr and SPIDEr-max scores, we computed the difference between them with the best candidate given by the model and the best one given by the SPIDEr score for various beam sizes.



The correlation value between this variation of TF-IDF and SPIDEr scores is almost one for AudioCaps and Clotho. It suggests that the candidates selected by SPIDEr-max have a much higher TF-IDF than those selected by the model probabilities, which appears to significantly increase the CIDEr-D score and, thus, also the SPIDEr-max score.

\section{Conclusion}
\label{sec:conclusion}
In this paper, we showed that the SPIDEr score is very sensitive to the words used in the caption candidates, so we proposed a new metric, SPIDEr-max, that takes into account multiple candidates for each audio recording. The scores of SPIDEr-max compared to human scores of SPIDEr show that our model already produces human-like caption candidates, but selecting the caption with the highest SPIDEr score is not trivial. As future work, we are interested to study other metrics that do not use TF-IDF, such as model-based metrics like BERTScore~\cite{zhang_bertscore_2020} or FENSE~\cite{zhou_can_2022}. We also look forward to testing SPIDEr-max with new models to see if our findings are repeated across architectures and training methods.




\section{ACKNOWLEDGMENT}

This work was partially supported by the Agence Nationale de la Recherche LUDAU (Lightly-supervised and Unsupervised Discovery of Audio Units using Deep Learning) project (ANR-18-CE23-0005-01). Experiments presented in this paper were carried out using the OSIRIM platform that is administered by IRIT and supported by CNRS, the Region Midi-Pyrénées, the French Government, ERDF (see http://osirim.irit.fr/). 

\bibliographystyle{IEEEtran}
\bibliography{refs}

\begin{thebibliography}{10}
\providecommand{\url}[1]{#1}
\def\UrlFont{\rmfamily}
\providecommand{\newblock}{\relax}
\providecommand{\bibinfo}[2]{#2}
\providecommand\BIBentrySTDinterwordspacing{\spaceskip=0pt\relax}
\providecommand\BIBentryALTinterwordstretchfactor{4}
\providecommand\BIBentryALTinterwordspacing{\spaceskip=\fontdimen2\font plus
\BIBentryALTinterwordstretchfactor\fontdimen3\font minus
  \fontdimen4\font\relax}
\providecommand\BIBforeignlanguage[2]{{%
\expandafter\ifx\csname l@#1\endcsname\relax
\typeout{** WARNING: IEEEtran.bst: No hyphenation pattern has been}%
\typeout{** loaded for the language `#1'. Using the pattern for}%
\typeout{** the default language instead.}%
\else
\language=\csname l@#1\endcsname
\fi
#2}}

\bibitem{xu2022comprehensive}
X.~Xu, M.~Wu, and K.~Yu, ``A comprehensive survey of automated audio
  captioning,'' \emph{arXiv preprint arXiv:2205.05357}, 2022.

\bibitem{kong_panns_2020}
Q.~Kong, Y.~Cao, T.~Iqbal, Y.~Wang, W.~Wang, and M.~D. Plumbley,
  ``\BIBforeignlanguage{en}{{PANNs}: {Large}-{Scale} {Pretrained} {Audio}
  {Neural} {Networks} for {Audio} {Pattern} {Recognition}},''
  \emph{\BIBforeignlanguage{en}{arXiv:1912.10211 [cs, eess]}}, Aug. 2020,
  arXiv: 1912.10211.

\bibitem{xinhao2021_t6}
X.~Mei, Q.~Huang, X.~Liu, G.~Chen, J.~Wu, Y.~Wu, J.~Zhao, S.~Li, T.~Ko, H.~L.
  Tang, X.~Shao, M.~D. Plumbley, and W.~Wang, ``An encoder-decoder based audio
  captioning system with transfer and reinforcement learning for {DCASE}
  challenge 2021 task 6,'' DCASE2021 Challenge, Tech. Rep., July 2021.

\bibitem{gontier:hal-03522488}
F.~Gontier, R.~Serizel, and C.~Cerisara, ``{Automated audio captioning by
  fine-tuning bart with audioset tags},'' in \emph{{DCASE 2021 - 6th Workshop
  on Detection and Classification of Acoustic Scenes and Events}}, Virtual,
  Spain, Nov. 2021.

\bibitem{liu_improved_2017}
S.~Liu, Z.~Zhu, N.~Ye, S.~Guadarrama, and K.~Murphy,
  ``\BIBforeignlanguage{en}{Improved {Image} {Captioning} via {Policy}
  {Gradient} optimization of {SPIDEr}},'' \emph{\BIBforeignlanguage{en}{2017
  IEEE International Conference on Computer Vision (ICCV)}}, pp. 873--881, Oct.
  2017, arXiv: 1612.00370.

\bibitem{vedantam_cider_2015}
R.~Vedantam, C.~L. Zitnick, and D.~Parikh, ``{CIDEr}: {Consensus}-based {Image}
  {Description} {Evaluation},'' \emph{arXiv:1411.5726 [cs]}, June 2015, arXiv:
  1411.5726.

\bibitem{anderson_spice_2016}
P.~Anderson, B.~Fernando, M.~Johnson, and S.~Gould,
  ``\BIBforeignlanguage{en}{{SPICE}: {Semantic} {Propositional} {Image}
  {Caption} {Evaluation}},'' \emph{\BIBforeignlanguage{en}{arXiv:1607.08822
  [cs]}}, July 2016, arXiv: 1607.08822.

\bibitem{klein_accurate_2003}
D.~Klein and C.~D. Manning, ``\BIBforeignlanguage{en}{Accurate unlexicalized
  parsing},'' in \emph{\BIBforeignlanguage{en}{Proceedings of the 41st {Annual}
  {Meeting} on {Association} for {Computational} {Linguistics} - {ACL} '03}},
  vol.~1.\hskip 1em plus 0.5em minus 0.4em\relax Sapporo, Japan: Association
  for Computational Linguistics, 2003, pp. 423--430.

\bibitem{hsu_text-free_2020}
W.-N. Hsu, D.~Harwath, C.~Song, and J.~Glass,
  ``\BIBforeignlanguage{en}{Text-{Free} {Image}-to-{Speech} {Synthesis} {Using}
  {Learned} {Segmental} {Units}},'' Dec. 2020, arXiv:2012.15454 [cs, eess].

\bibitem{papineni_bleu_2001}
K.~Papineni, S.~Roukos, T.~Ward, and W.-J. Zhu,
  ``\BIBforeignlanguage{en}{{BLEU}: a method for automatic evaluation of
  machine translation},'' in \emph{\BIBforeignlanguage{en}{Proceedings of the
  40th {Annual} {Meeting} on {Association} for {Computational} {Linguistics} -
  {ACL} '02}}.\hskip 1em plus 0.5em minus 0.4em\relax Philadelphia,
  Pennsylvania: Association for Computational Linguistics, 2001, p. 311.

\bibitem{lin-2004-rouge}
C.-Y. Lin, ``{ROUGE}: A package for automatic evaluation of summaries,'' in
  \emph{Text Summarization Branches Out}.\hskip 1em plus 0.5em minus
  0.4em\relax Barcelona, Spain: Association for Computational Linguistics, July
  2004, pp. 74--81.

\bibitem{denkowski_meteor_2014}
M.~Denkowski and A.~Lavie, ``\BIBforeignlanguage{en}{Meteor {Universal}:
  {Language} {Specific} {Translation} {Evaluation} for {Any} {Target}
  {Language}},'' in \emph{\BIBforeignlanguage{en}{Proceedings of the {Ninth}
  {Workshop} on {Statistical} {Machine} {Translation}}}.\hskip 1em plus 0.5em
  minus 0.4em\relax Baltimore, Maryland, USA: Association for Computational
  Linguistics, 2014, pp. 376--380.

\bibitem{drossos_clotho_2019}
K.~Drossos, S.~Lipping, and T.~Virtanen, ``\BIBforeignlanguage{en}{Clotho: {An}
  {Audio} {Captioning} {Dataset}},''
  \emph{\BIBforeignlanguage{en}{arXiv:1910.09387 [cs, eess]}}, Oct. 2019,
  arXiv: 1910.09387.

\bibitem{kim-etal-2019-audiocaps}
C.~D. Kim, B.~Kim, H.~Lee, and G.~Kim, ``{A}udio{C}aps: Generating captions for
  audios in the wild,'' in \emph{Proceedings of the 2019 Conference of the
  North {A}merican Chapter of the Association for Computational Linguistics:
  Human Language Technologies, Volume 1 (Long and Short Papers)}.\hskip 1em
  plus 0.5em minus 0.4em\relax Minneapolis, Minnesota: Association for
  Computational Linguistics, June 2019, pp. 119--132.

\bibitem{gemmeke_audio_2017}
J.~F. Gemmeke, D.~P.~W. Ellis, D.~Freedman, A.~Jansen, W.~Lawrence, R.~C.
  Moore, M.~Plakal, and M.~Ritter, ``\BIBforeignlanguage{en}{Audio {Set}: {An}
  ontology and human-labeled dataset for audio events},'' in
  \emph{\BIBforeignlanguage{en}{2017 {IEEE} {International} {Conference} on
  {Acoustics}, {Speech} and {Signal} {Processing} ({ICASSP})}}.\hskip 1em plus
  0.5em minus 0.4em\relax New Orleans, LA: IEEE, Mar. 2017, pp. 776--780.

\bibitem{ines_montani_2021_5764736}
I.~Montani, M.~Honnibal, M.~Honnibal, S.~V. Landeghem, A.~Boyd, H.~Peters,
  P.~O. McCann, M.~Samsonov, J.~Geovedi, J.~O'Regan, G.~Orosz, D.~Altinok,
  S.~L. Kristiansen, Roman, E.~Bot, L.~Fiedler, G.~Howard, W.~Phatthiyaphaibun,
  Y.~Tamura, S.~Bozek, murat, M.~Amery, B.~Böing, P.~K. Tippa, L.~U.
  Vogelsang, B.~Vanroy, R.~Balakrishnan, V.~Mazaev, and GregDubbin,
  ``{explosion/spaCy: v3.2.1: doc\_cleaner component, new Matcher attributes,
  bug fixes and more},'' Dec. 2021.

\bibitem{DBLP:journals/corr/VaswaniSPUJGKP17}
A.~Vaswani, N.~Shazeer, N.~Parmar, J.~Uszkoreit, L.~Jones, A.~N. Gomez,
  L.~Kaiser, and I.~Polosukhin, ``Attention is all you need,'' \emph{CoRR},
  vol. abs/1706.03762, 2017.

\bibitem{kingma_adam_2017}
D.~P. Kingma and J.~Ba, ``\BIBforeignlanguage{en}{Adam: {A} {Method} for
  {Stochastic} {Optimization}},'' \emph{\BIBforeignlanguage{en}{arXiv:1412.6980
  [cs]}}, Jan. 2017, arXiv: 1412.6980.

\bibitem{park19e_interspeech}
D.~S. Park, W.~Chan, Y.~Zhang, C.-C. Chiu, B.~Zoph, E.~D. Cubuk, and Q.~V. Le,
  ``{SpecAugment: A Simple Data Augmentation Method for Automatic Speech
  Recognition},'' in \emph{Proc. Interspeech 2019}, 2019, pp. 2613--2617.

\bibitem{NIPS2019_9015}
A.~Paszke, S.~Gross, F.~Massa, A.~Lerer, J.~Bradbury, G.~Chanan, T.~Killeen,
  Z.~Lin, N.~Gimelshein, L.~Antiga, A.~Desmaison, A.~Kopf, E.~Yang, Z.~DeVito,
  M.~Raison, A.~Tejani, S.~Chilamkurthy, B.~Steiner, L.~Fang, J.~Bai, and
  S.~Chintala, ``Pytorch: An imperative style, high-performance deep learning
  library,'' in \emph{proc. NeurIPS}, 2019, pp. 8026--8037.

\bibitem{falcon2019pytorch}
W.~Falcon and .al, ``Pytorch lightning,'' \emph{GitHub. Note:
  https://github.com/PyTorchLightning/pytorch-lightning}, vol.~3, 2019.

\bibitem{xu2022_t6a}
X.~Xu, Z.~Xie, M.~Wu, and K.~Yu, ``The {SJTU} system for {DCASE2022} challenge
  task 6: Audio captioning with audio-text retrieval pre-training,'' DCASE2022
  Challenge, Tech. Rep., July 2022.

\bibitem{zhang_bertscore_2020}
T.~Zhang, V.~Kishore, F.~Wu, K.~Q. Weinberger, and Y.~Artzi,
  ``\BIBforeignlanguage{en}{{BERTScore}: {Evaluating} {Text} {Generation} with
  {BERT}},'' \emph{\BIBforeignlanguage{en}{arXiv:1904.09675 [cs]}}, Feb. 2020,
  arXiv: 1904.09675.

\bibitem{zhou_can_2022}
Z.~Zhou, Z.~Zhang, X.~Xu, Z.~Xie, M.~Wu, and K.~Q. Zhu,
  ``\BIBforeignlanguage{en}{Can {Audio} {Captions} {Be} {Evaluated} with
  {Image} {Caption} {Metrics}?}'' Jan. 2022, number: arXiv:2110.04684
  arXiv:2110.04684 [cs, eess].

\end{thebibliography}

\end{sloppy}
\end{document}